\documentclass[fleqn,usenatbib]{mnras}
\usepackage{amssymb,amsfonts,amsmath,bm}
\usepackage{graphicx}
\usepackage{color}
\usepackage{mathrsfs}
\usepackage{hyperref}

\DeclareMathOperator{\arctanh}{arctanh}

\title[Mass and deformation effects on periapsis shift]{Extended mass and spheroidal deformation effects on epicyclic frequencies and periapsis shift in quasi-circular orbits
}

\author[T. Igata]{
Takahisa Igata\thanks{
E-mail: takahisa.igata@gakushuin.ac.jp
} \\
Department of Physics, Gakushuin University, Mejiro, Toshima, Tokyo 171-8588, Japan
}

\begin{document}
\maketitle

\begin{abstract}
We investigate the effects of extended mass and spheroidal deformation on the periapsis shift of quasi-circular orbits inside a gravitating mass distribution in the Newtonian framework. The analysis is restricted to orbits confined to the reflection-symmetric plane of the spheroidal configuration. Focusing on the internal gravitational potential of a spheroidal body with both homogeneous and inhomogeneous density profiles, we elucidate how the ratio of local density to average density governs the extended mass effect on the periapsis shift. By analysing the orbital angular frequency, along with the radial and vertical epicyclic frequencies, we demonstrate that in the uniform density case (i.e., the Maclaurin spheroid), where the potential takes the form of a harmonic oscillator, the periapsis exhibits a constant retrograde shift of $-\pi$. In contrast, in regions where density inhomogeneity and spheroidal deformation (in both prolate and oblate forms) are significant, the periapsis shift varies with the guiding orbital radius due to local density contrast and deformation effects. The results indicate that oblate deformation suppresses the extended mass effect associated with the ratio of local density to average density, whereas prolate deformation amplifies it. Furthermore, by varying the density distribution parameters, we establish the conditions for orbital stability and identify the emergence of marginally stable orbits.
\end{abstract}


\section{Introduction}
\label{sec:1}
Quasi-circular orbits, which are derived by applying small perturbations to stable circular orbits, provide a fundamental framework in astrophysics due to their straightforward geometric nature. The decomposition into a primary circular motion and superposed radial harmonic (epicyclic) oscillations enables a clear separation of physical effects, since the difference between the radial epicyclic frequency and the orbital angular frequency results in a periapsis shift. This characteristic makes quasi-circular orbits an ideal probe for investigating the local properties of gravitational fields.

Recent advances in observations of the Galactic center have provided detailed insights into the gravitational field surrounding the supermassive object~\citep{Ghez:2000ay,Schodel:2002py}. For example, precise tracking of S2's (also known as S0-2) orbit confirms a subtle prograde periapsis shift that is consistent with the predictions of general relativity~\citep{GRAVITY:2018ofz,Do:2019txf,Saida:2019mcz}. Furthermore, higher-order relativistic cross-term effects in the perihelion precession of Mercury may be detectable by the BepiColombo mission~\citep{Will:2018mcj,BepiColombo:2021}.

Motivated by these findings, theoretical research in this field has grown increasingly active. Various approaches have been developed to address periapsis shifts, including the post-Newtonian approximation~\citep{Damour:1988mr,Damour:2000we,Yamada:2012xt,PoissonWill:2014,Iwata:2016ivt}, the numerical relativity~\citep{Mroue:2010re,LeTiec:2011bk}, and series expansion methods in gravitational theory~\citep{Bini:2005dy,Vogt:2008zs,Mak:2018hcn,Tucker:2018rgy,He:2023joa}. Moreover, a novel expansion method has recently been introduced~\citep{Walters:2018,Katsumata:2024qzv}. Since quasi-circular orbits constitute the leading order in the eccentricity expansion, a thorough understanding of them is essential for accurately determining higher-order corrections and periapsis shifts.

Consequently, our understanding of periapsis shifts has significantly advanced. Furthermore, in realistic astrophysical environments, an additional factor---the extended mass effect---must be considered. In the ideal Kepler problem, bound orbits are perfectly elliptical and exhibit no periapsis shift. In contrast, when a gravitational source possesses an extended mass distribution and a test particle moves through regions of non-zero mass density, the orbital dynamics are fundamentally altered. In such cases, periapsis shifts originate from a mechanism distinct from that of relativistic corrections, and they exhibit markedly different characteristics.

For a uniform-density sphere, a test particle moving through only part of the interior experiences a retrograde periapsis shift~\citep{1985AmJPh..53..694J}, whereas an orbit entirely within the sphere---reflecting a harmonic oscillator potential---results in a retrograde shift of $-\pi$ (see, e.g., \citealt{BinneyTremaine:2008}). Inhomogeneous density distributions further reveal a competition between general relativistic corrections and Newtonian extended mass effects~\citep{Rubilar:2001}. Furthermore, models incorporating relativistic corrections to the extended mass effects have been applied to scenarios including black holes with matter fields~\citep{DellaMonica:2021xcf,Bambhaniya:2025xmu}, dark matter cores~\citep{2022MNRAS.511L..35A,Igata:2022nkt}, black holes with cloud systems~\citep{Sadeghian:2013laa,Igata:2022rcm, Jusufi:2022jxu}, boson stars~\citep{Grould:2017rzz}, wormhole~\citep{Manna:2019,DeFalco:2021btn,Deligianni:2021hwt}, and naked singularities~\citep{Stefanov:2012fb,Bambhaniya:2021ybs,Bambhaniya:2022xbz,Ota:2021mub,Turimov:2024oxn}. 

Recently, a general formula for periapsis shifts in static, spherically symmetric spacetimes was proposed, providing a unified framework that transcends previous case studies~\citep{Harada:2022uae}. In this approach, the formula is expressed in terms of the invariant local curvature, the guiding orbital radius, and the quasi-local mass [see Eq.~\eqref{eq:DphipGR}], which renders it applicable to a variety of gravitational theories, including general relativity. Since curvature directly reflects the energy density and pressure of matter fields through the field equations, this formula is essential for interpreting extended mass effects and deepening our understanding of the phenomenon.

Realistic astrophysical mass distributions frequently deviate from perfect spherical symmetry, thereby significantly affecting the orbital dynamics.%
\footnote{For example, the subtle deformation of the Sun and its impact on the periapsis shift in the Solar system have been extensively analysed using the post-Newtonian method~\citep{PoissonWill:2014}.} Recent studies have investigated epicyclic oscillations in Newtonian gravitational fields generated by ellipsoidal mass distributions. In the static, axisymmetric vacuum field of an oblate spheroid with uniform density (i.e., a Maclaurin spheroid), the epicyclic frequencies have been studied in detail~\citep{Amsterdamski:2000mq,Kluzniak:2013}, revealing a striking similarity to those calculated for the Kerr spacetime in \citet{Schmidt:2002qk}. In this setting, when the oblateness exceeds a critical value, the radial epicyclic frequency of quasi-circular orbits on the equatorial plane reaches a maximum at a specific radius and subsequently vanishes near the spheroid's surface, thereby inducing orbital instability. Such behaviours arise from contributions of higher-order mass multipoles~\citep{Shibata:1998xw,Zdunik:2000bc}. Moreover, analyses have been extended to ellipsoidal models with inhomogeneous density, further including prolate configurations: in oblate models with a relatively gradual central density profile, an unstable region for quasi-circular orbits emerges near the surface (similar to uniform models), whereas oblate models with a steeper central density concentration stabilize the orbits; in prolate models, the radial epicyclic frequency increases monotonically as the guiding orbital radius decreases, thereby ensuring stability over the entire region~\citep{Bollimpalli:2022xgw}.

Although previous studies have clarified the deformation effects on epicyclic oscillations in external vacuum fields, such effects remain underexplored in mass distributions deviating from spherical symmetry. In scenarios where local matter fields are significant, the periapsis shift may deviate from predictions based solely on relativistic corrections. Addressing this gap is essential for developing a unified formula that fully incorporates general relativistic effects and implies that deformed environments can induce novel phenomena even within the framework of Newtonian mechanics.

Motivated by this background, we examine the periapsis shift in quasi-circular orbits within Newtonian theory by employing a generalized mass distribution model that extends the conventional spherically symmetric case by incorporating a one-parameter deformation. Throughout this work, we restrict our analysis to quasi-circular orbits that lie in the reflection-symmetric (equatorial) plane of the mass distribution. We investigate how this deformation influences extended mass effects by establishing a quantitative relation between the periapsis shift angle and both local and quasi-local average mass densities.

This paper is organized as follows. In Sec.~\ref{sec:2}, we review extended mass effects on the periapsis shift, focusing on the local-to-average density ratio in a static, spherically symmetric model. In Sec.~\ref{sec:3}, we derive the gravitational potential for an ellipsoidal mass distribution, focusing on a solid spheroid. In Sec.~\ref{sec:4}, we formulate the motion of a test particle on quasi-circular orbits within a reflection symmetric plane, deriving the orbital angular and epicyclic frequencies, and defining the periapsis shift angle. In Section~\ref{sec:5}, we analyse the frequency and periapsis shift properties for various density distributions and discuss the effects of source deformation. Finally, in Sec.~\ref{sec:6}, we summarize our results and outline directions for future research.

\section{Extended mass effects on the periapsis shift of quasi-circular orbits in a spherically symmetric Newtonian system}
\label{sec:2}
In this section, we examine the periapsis shift angle of quasi-circular orbits in a static, spherically symmetric Newtonian gravitational system, and review the extended mass effects contributing to this shift.

First, we consider a quasi-circular orbit, defined as an orbit with a negligibly small eccentricity achieved by applying a small radial perturbation to an otherwise stable circular orbit. In other words, a quasi-circular orbit oscillates harmonically in the radial direction at an epicyclic frequency $\omega_{\mathrm{r}}$, with its motion centered on the stable circular guiding orbit. Furthermore, a periapsis shift occurs when the ratio of the orbital angular frequency $\omega_{\phi}$ to the radial oscillation frequency $\omega_{\mathrm{r}}$ deviates from unity. The periapsis shift angle, which quantifies this effect, is defined as
\begin{align}
\Delta \phi_{\mathrm{p}}=2\pi \left(
\frac{\omega_\phi}{\omega_{\mathrm{r}}}-1
\right).
\label{eq:Deltaphip0}
\end{align}

Recently, a general formula for the periapsis shift angle of quasi-circular orbits of a test particle in a static and spherically symmetric spacetime was presented in \citet{Harada:2022uae} as
\begin{align}
 \Delta\phi_{\mathrm{p}}={}&2\pi \left\{ \left[\:\!
1-\frac{6m(r)}{r}+4\pi r^{2}
\left[\:\! \rho(r)-3P(r)\:\!\right]\right.\right.
\nonumber
\\
&
\!+\bigg(1-\frac{2m(r)}{r}\bigg)\frac{ \rho(r)+P(r)+2\Pi(r)}{\frac{m(r)}{4\pi r^{3}}+P(r)}
\:\!\bigg]^{-1/2}\!-1
\bigg\},
\label{eq:DphipGR}
\end{align}
where we use geometrized units in which $G=1$ and $c=1$. Here, $r$ denotes the radius of the orbit's guiding center, $m(r)$ is the quasi-local mass contained within the sphere of radius $r$, and $\rho(r)$, $P(r)$, and $\Pi(r)$ represent the local mass density, radial pressure, and tangential pressure of matter fields, respectively. In the Newtonian limit, the formula reduces to the following expression for a static and spherically symmetric gravitational system:
\begin{align}
\Delta \phi_{\mathrm{p}}=2\pi \left[\:\!
\left(1+\frac{3\rho(r)}{\bar{\rho}(r)}\right)^{-1/2}-1
\:\!\right],
\label{eq:Newtonlimit}
\end{align}
where $\bar{\rho}(r)=3m/(4\pi r^3)$ denotes the average mass density within the radius $r$. It should be noted that this expression incorporates extended mass effects. Specifically, the ratio $\rho(r)/\bar{\rho}(r)$ reflects the balance between the local and quasi-local contributions of the distributed mass to the periapsis shift angle, thereby distinguishing it from the point-mass case.

\subsection{Diffuse regime: $\rho/\bar{\rho}\ll 1$}
We focus on the case where the local density is significantly smaller than the average density around the guiding center, i.e., $\rho/\bar{\rho}\ll 1$, which we refer to as the diffuse regime. In this regime, the periapsis shift angle is given by 
\begin{align}
\Delta \phi_{\mathrm{p}}=-3\pi \frac{\rho}{\bar{\rho}}+O\left(\Big(\frac{\rho}{\bar{\rho}}\Big)^2\right).
\end{align}
This expression shows that the constant term vanishes, implying that no periapsis shift occurs if the orbital region is locally vacuum (i.e., $\rho(r)=0$). Consequently, the periapsis shift is only slightly negative for $0<\rho/\bar{\rho}\ll1$, indicating that extended mass effects contribute negatively to the periapsis shift.

\subsection{Homogeneous regime: $\rho/\bar{\rho}\approx 1$}
We consider the case where the local density is comparable to the average density around the guiding center, i.e., $\rho/\bar{\rho}\approx 1$, which we refer to as the homogeneous regime. In this regime, the periapsis shift angle is given by
\begin{align}
\Delta \phi_{\mathrm{p}} = -\pi - \frac{3\pi}{8}\left(\frac{\rho}{\bar{\rho}} - 1\right)+O\left(\Big(\frac{\rho}{\bar{\rho}} - 1\Big)^2\right),
\label{eq:Hregime}
\end{align}
which implies that $\Delta \phi_{\mathrm{p}}$ is approximately $-\pi$. In the case of a uniform density sphere, where $\rho=\bar{\rho}$ holds exactly, we obtain the result $\Delta \phi_{\mathrm{p}}=-\pi$.

\subsection{Spike regime: $\rho/\bar{\rho}\gg 1$}
We consider the case where the local density significantly exceeds the average density around the guiding center, i.e., $\rho/\bar{\rho}\gg 1$, which we refer to as the spike regime. In this regime, the periapsis shift angle is given by
\begin{align}
\Delta \phi_{\mathrm{p}} = -2\pi+2\pi\sqrt{\frac{\bar{\rho}}{3\rho}}+O\left(\Big(\frac{\bar{\rho}}{\rho}\Big)^{3/2}\right).
\end{align}
Thus, $\Delta \phi_{\mathrm{p}}$ is approximately $-2\pi$, indicating that a high local density gives rise to dominant radial oscillations around the guiding orbit.

\section{Gravitational potential of an inhomogeneous spheroid}
\label{sec:3}

In this section, we review the gravitational potential inside an inhomogeneous spheroid~\citep{Chandrasekhar:1969}. Hereafter, we work in the Newtonian framework and retain the gravitational constant $G$ explicitly, in contrast to the use of geometrized units in Sec.~\ref{sec:2}. Let $x_i$ ($i=1, 2, 3$) denote the Cartesian coordinates in the 3D Euclidean space $\mathbb{R}^3$. We consider a solid ellipsoid with semi-axis lengths $a_1$, $a_2$, and $a_3$, where ``solid" refers to a mass distribution that fills the volume, not to mechanical rigidity. We assume that the solid ellipsoid is foliated by thin, homogeneous homeoids, each with semi-axis lengths given by $m a_i$ ($i=1, 2, 3$), where the parameter $m$ varies from $0$ at the center to $1$ at the surface. Each homeoid is defined by
\begin{align}
\sum_{i=1}^3\frac{x_i^2}{a_i^2}=m^2 \ \ (0\le m \le 1).
\end{align}
This result implies that the density of this ellipsoid depends solely on $m^2$, i.e., $\rho=\rho(m^2)$. Such an inhomogeneous ellipsoidal mass distribution yields an interior gravitational potential (see Theorem~12 in \citet{Chandrasekhar:1969}), 
\begin{align}
\Phi=-\pi \:\!G a_1 a_2 a_3 
\int_0^\infty \frac{\mathrm{d}u}{\Delta}
\int_{m^2(u)}^1 \mathrm{d}m^2 \:\!\rho(m^2),
\label{eq:Phigeneral}
\end{align}
where $G$ denotes the gravitational constant, and the functions $\Delta$ and $m^2(u)$ are defined as follows: 
\begin{align}
m^2(u)&=\sum_{i=1}^3 \frac{x_i^2}{a_i^2+u}
\\
\Delta&=\left[\:\!(a_1^2+u)(a_2^2+u)(a_3^2+u)\:\!\right]^{1/2}.
\end{align}

Now we assume the following form for $\rho(m^2)$:
\begin{align}
\rho(m^2)=\left\{
\begin{array}{ll}
\rho_0\left(1-m^2\right)^n
&\ \ (m\le 1)
\\
0&\ \ (m>1), 
\end{array}
\right.
\label{eq:rho}
\end{align}
where $\rho_0$ is a constant, and $n$ is a positive number. When $n=0$, Eq.~\eqref{eq:rho} reduces to a constant density within the spheroid, corresponding to the Maclaurin spheroid. This density profile corresponds to the Ferrers model, which is widely used in galactic dynamics to represent ellipsoidal mass distributions such as galactic bars and bulges~\citep{BinneyTremaine:2008}. Although it does not describe a hydrostatic equilibrium configuration as in polytropic models, nor does it share the extended asymptotic tail of the Plummer profile, it captures the essential feature of central mass concentration. The index $n$ controls the steepness of this concentration, allowing for systematic variation of the density profile. We adopt this model primarily for its analytical tractability, which enables exact evaluation of the internal gravitational potential and facilitates investigation of deformation effects. Substituting Eq.~\eqref{eq:rho} into Eq.~\eqref{eq:Phigeneral} and performing the integration, we obtain 
\begin{align}
\Phi=
-\pi \:\!G a_1 a_2 a_3 \frac{\rho_0}{n+1}
\int_0^\infty 
\frac{\mathrm{d}u}{\Delta}\left(
1-\sum_{i=1}^3\frac{x_i^2}{a_i^2+u}
\right)^{n+1},
\label{eq:Phintriaxial}
\end{align}
which is referred to as the Ferrers potential.

In the following, we focus on spheroids, i.e., 
\begin{align}
a_1=a_2=a, \quad a_3=b.
\end{align}
The residual axial symmetry of spheroids simplifies the analysis of their properties. For an inhomogeneous spheroid, the potential~\eqref{eq:Phintriaxial} reduces to
\begin{align}
\Phi_n(R, z)=&-\pi \:\!G \:\!a^2 b\:\!\frac{\rho_0}{n+1}
\int_0^\infty
\frac{\mathrm{d}u}{\Delta}
\nonumber
\\
&\times\left(
1-\frac{R^2}{a^2+u}-\frac{z^2}{b^2+u}
\right)^{n+1},
\end{align}
where $R=(x_1^2+x_2^2)^{1/2}$, $z=x_3$, and $\tan \phi=x_2/x_1$ denote the cylindrical coordinates. If $n$ is a non-negative integer, then the potential $\Phi_n$ can be expressed in the following systematic form:
\begin{align}
\Phi_n(R, z)={}&-n! \:\!\pi \:\!G \:\!\rho_0 \:\!a^2 
\nonumber
\\
&\times\sum_{\substack{\alpha+\beta+\gamma=n+1\\\alpha, \beta, \gamma\ge 0}}
(-1)^{\alpha+\beta}\frac{C_{\alpha, \beta}}{\alpha!\:\!\beta!\:\!\gamma!} 
\Big(\frac{R}{a}\Big)^{2\alpha}
\Big(\frac{z}{b}\Big)^{2\beta},\!\!
\label{eq:Phiexp}
\end{align}
where $C_{\alpha, \beta}$ are positive definite, dimensionless parameters given by
\begin{align}
C_{\alpha, \beta}=\int_0^\infty \frac{a^{2\alpha} b^{2\beta+1}}{(a^2+u)^{\alpha+1}(b^2+u)^{\beta+1/2}}\mathrm{d}u.
\label{eq:C}
\end{align}
For oblate spheroids (i.e., $a>b$), we define the oblateness as $e=\sqrt{1-b^2/a^2}$, while for prolate spheroids (i.e., $a<b$), we define the prolateness as $e=\sqrt{1-a^2/b^2}$. Consequently, we can express $C_{\alpha,\beta}$ as functions of $e$. Explicit forms of $C_{\alpha,\beta}$ are summarized in Appendix~\ref{sec:A}.

We focus on a spheroid whose boundary is defined by the constant level surface $m^2=\mathrm{constant}$. The mass of the spheroid is obtained by performing the following integration:
\begin{align}
M_n(m^2)=\int_0^{m^2}\rho(m^2) \frac{\mathrm{d}V}{\mathrm{d}m^2}\:\!\mathrm{d}m^2.
\end{align}
Since the volume enclosed by the spheroidal layer corresponding to $m^2$ is given by $V(m^2)=\frac{4}{3}\pi a^2 b m^3$, 
the evaluation of the integral yields
\begin{align}
M_n(m^2)=2\pi \rho_0 a^2 b B_{m^2} \bigg(\frac{3}{2}, 1+n\bigg),
\end{align}
where $B_{x}(\alpha, \beta)$ denotes the incomplete beta function defined by
\begin{align}
B_{x}(\alpha, \beta)\equiv \int_0^x t^{\alpha-1} (1-t)^{\beta-1} \:\!\mathrm{d}t.
\end{align}
Furthermore, the average density within the spheroidal layer specified by $m^2$ is defined as 
\begin{align}
\bar{\rho}(m^2):=\frac{M_n(m^2)}{V(m^2)}.
\end{align}
Upon evaluation, we obtain
\begin{align}
\bar{\rho}(m^2)= \frac{3\rho_0B_{m^2}\big(\frac{3}{2}, 1+n\big)}{2m^3}.
\label{eq:barrho}
\end{align}

\section{Epicyclic oscillations and periapsis shift in quasi-circular orbits}
\label{sec:4}
We consider the motion of a particle in the gravitational potential described by Eq.~\eqref{eq:Phiexp}. The Lagrangian per unit rest mass is given by
\begin{align}
\mathscr{L}=\frac{1}{2}(\dot{R}^2+R^2\dot{\phi}^2+\dot{z}^2
)-\Phi_n(R,z),
\end{align}
where the dot denotes differentiation with respect to time. Since the coordinate $\phi$ is cyclic, 
the Euler--Lagrange equation for $\phi$ implies the conservation of angular momentum, yielding $R^2 \dot{\phi}=l$, where $l$ denotes the constant specific angular momentum. This conservation of angular momentum reduces the 3D motion to an effective 2D motion in the $R$--$z$ plane, which is described by the Hamiltonian
\begin{align}
H_{\mathrm{eff}}=\frac{1}{2}\left(p_{R}^2+p_z^2\right)+\Phi_{n, \mathrm{eff}},
\end{align}
where $p_{R}$ and $p_z$ denote the conjugate momenta to $R$ and $z$, respectively, and 
\begin{align}
\Phi_{n, \mathrm{eff}}=\frac{l^2}{2R^2}+\Phi_n(R,z),
\label{eq:Phieff}
\end{align}
which defines the effective potential. The first term on the right-hand side corresponds to the centrifugal potential.

We focus on circular orbits confined to the equatorial plane (i.e., $z=0$). Note that, due to the reflection symmetry ($z\to -z$) of the system, if both the initial position and velocity are confined to the equatorial plane, the orbit remains in that plane. A necessary condition for $\Phi_{n,\mathrm{eff}}$ to support circular orbits is the existence of a radius $R$ at which $\partial \Phi_{n,\mathrm{eff}}(R,0)/\partial R=0$. Solving this equation for $l^2$ yields
\begin{align}
l^2=R^3 \frac{\partial\Phi_n(R,0)}{\partial R}.
\label{eq:lsq}
\end{align}
Then, by invoking the conservation of angular momentum, the orbital angular frequency $\omega_{\phi}=\dot{\phi}$ for a particle in a circular orbit is given by
\begin{align}
\omega_{\phi}^2
&=\frac{1}{R} \left.\frac{\partial \Phi_n}{\partial R}\right|_{z=0}
\label{eq:omegaphi1}
\\
&=2\pi \:\!G\rho_0 \sum_{j=0}^n(-1)^{j}
\binom{n}{j}
C_{j+1, 0}\:\! \xi^{2j}. 
\label{eq:omegaphi2}
\end{align}
Throughout this paper, we define the dimensionless quantity $R/a$ as 
\begin{align}
\xi\equiv \frac{R}{a}.
\label{eq:xi}
\end{align}

Consider a small perturbation applied to a particle on a circular orbit, while preserving its angular momentum~\eqref{eq:lsq}. If the original circular orbit is stable, the small displacement will oscillate harmonically about the unperturbed orbit, which serves as the guiding center. In the linear perturbation regime, the radial and vertical oscillations, known as epicyclic oscillations, can be treated independently. The radial epicyclic frequency is defined by 
\begin{align}
\omega_{\mathrm{R}}^2
&= \left.\frac{\partial^2 \Phi_{n, \mathrm{eff}}}{\partial R^2}\right|_{z=0},
\label{eq:omegaR1}
\end{align}
and is given explicitly by
\begin{align}
\omega_{\mathrm{R}}^2
&=
4\pi \:\!G\rho_0 \sum_{j=0}^n(-1)^{j} (j+2)\binom{n}{j}
C_{j+1,0}\:\! \xi^{2j}.
\label{eq:omegaR2}
\end{align}
Similarly, the vertical epicyclic frequency is defined by
\begin{align}
\omega_{\mathrm{z}}^2
=\left.
\frac{\partial^2\Phi_{n, \mathrm{eff}}}{\partial z^2}
\right|_{z=0},
\label{eq:omegaz1}
\end{align}
which can be rewritten as
\begin{align}
\omega_{\mathrm{z}}^2=
2\pi \:\!G\rho_0 (a/b)^2 \sum_{j=0}^{n}(-1)^{j}
\binom{n}{j}
C_{j,1}\:\! \xi^{2j}.
\label{eq:omegaz2}
\end{align}
It should be noted that the necessary and sufficient condition for epicyclic oscillations is that the squared frequencies given in Eqs.~\eqref{eq:omegaphi1}, \eqref{eq:omegaR1}, and \eqref{eq:omegaz1} are all positive. These three squared frequencies and the density given in Eq.~\eqref{eq:rho} satisfy the relation
\begin{align}
\omega_{\mathrm{R}}^2+\omega_{\mathrm{z}}^2-2\omega_{\phi}^2=4\pi \:\!G \rho|_{z=0},
\end{align}
which is derived directly from the Poisson equation for a static, axisymmetric gravitational field~\citep{Kluzniak:2013,Vieira:2017zau,Delgado:2022yvg}.

If only radial oscillations are excited in a particle on a circular orbit (i.e., $\omega_{\mathrm{R}}^2\ge 0$), its orbit becomes quasi-circular in the equatorial plane. In general, the periapsis (or apoapsis) of such an orbit shifts continuously throughout the orbital motion. The periapsis shift angle is defined as
\begin{align}
\Delta \phi_{\mathrm{p}}=2\pi \left(
\frac{\omega_\phi}{\omega_{\mathrm{R}}}-1
\right).
\label{eq:Deltaphip}
\end{align}
One property of $\Delta \phi_{\mathrm{p}}$, independent of $n$, is that as $\xi \to 0$, the periapsis shift angle approaches $-\pi$. In the next section, we examine the properties of the orbital angular frequency, epicyclic frequencies, and periapsis shift for lower indices of $n$. 

By analogy with Eq.~\eqref{eq:Newtonlimit} for the spherically symmetric case, we assume that the periapsis shift angle, defined in Eq.~\eqref{eq:Deltaphip}, is expressed in the following form for a spheroidal mass distribution:
\begin{align}
\Delta \phi_{\mathrm{p}}=2\pi \left[\left(1+\frac{3\rho(\xi^2)}{\bar{\rho}(\xi^2)}+f\right)^{-1/2}-1 \right].
\label{eq:fdef}
\end{align}
Here, $\rho(\xi^2)$ and $\bar{\rho}(\xi^2)$, defined in Eqs.~\eqref{eq:rho} and \eqref{eq:barrho}, respectively, are both evaluated at $z=0$. Since $m^2=\xi^2$ at $z=0$, we neglect variations along the $z$-direction (e.g., due to deformation or density inhomogeneity), resulting in the same form as the spherically symmetric case. We introduce the phenomenological parameter $f$ to account for the additional contributions to the periapsis shift arising from deformation and associated density inhomogeneities, which are absent in a purely spherically symmetric system. When $f>0$, these influences not only amplify the extended mass effect but also add a purely deformation-induced contribution that further shifts the periapsis in the negative direction, yielding a more pronounced negative shift than that produced by the density ratio alone. Conversely, when $f<0$, the deformation effects are partially canceled, thereby reducing the magnitude of the negative periapsis shift. Thus, even in regions where the local density may vanish, these effects can lead to a non-zero value of $f$. In the spherically symmetric case (i.e., $e=0$), $f$ vanishes identically, thus recovering Eq.~\eqref{eq:Newtonlimit}.

\section{Properties of the epicyclic frequencies and the periapsis shift angle}
\label{sec:5}
\subsection{$n=0$}
\label{sec:5-0}
\begin{figure}
\centering
\includegraphics[width=\columnwidth]{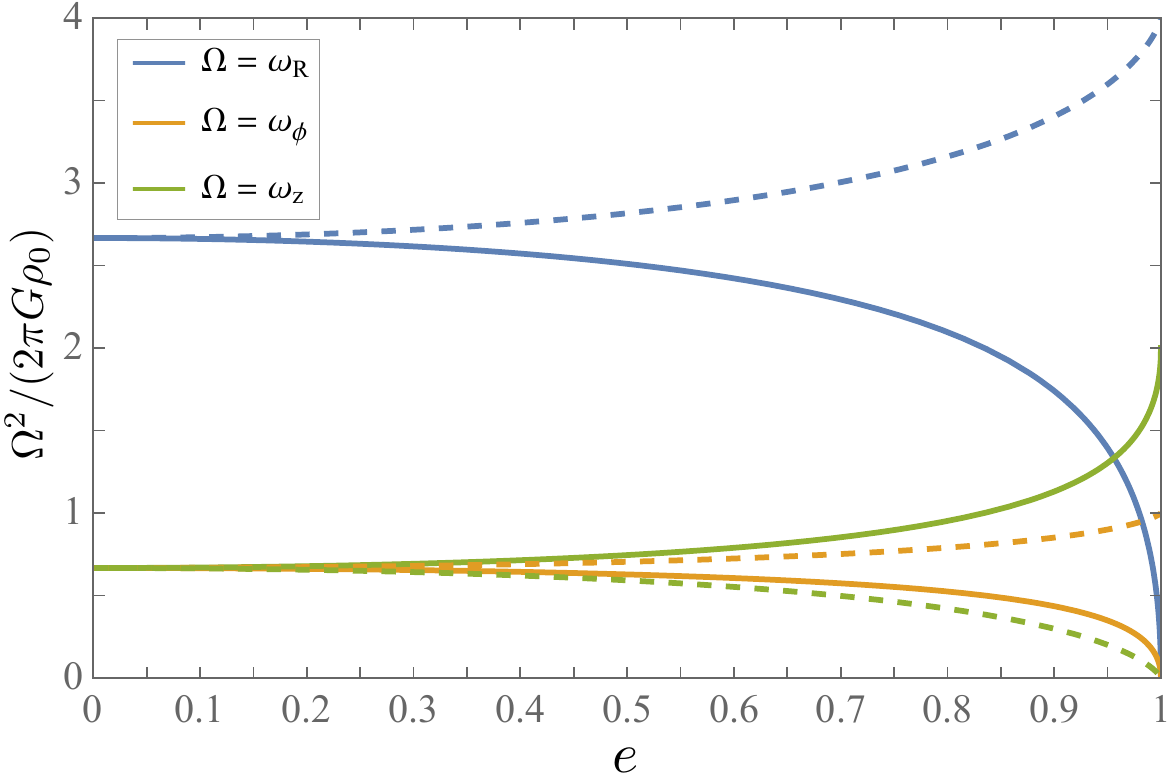}
\caption{Plots of the squares of the radial epicyclic frequency, orbital angular frequency, and vertical epicyclic frequency as functions of $e$ for an oblate spheroid (solid curves) and a prolate spheroid (dashed curves) with $n=0$.}
\label{fig:n0}
\end{figure}

We begin by considering the case $n=0$, corresponding to a spheroid with a uniform density distribution (the Maclaurin spheroid). Inside the spheroid, the particle dynamics is governed by the effective potential~\eqref{eq:Phieff} (with $n=0$), which is given explicitly by 
\begin{align}
\Phi_{0, \mathrm{eff}}=\frac{l^2}{2R^2}-\pi \:\!G \rho_0 a^2\left[\:\!
C_{0,0}-C_{1,0}\Big(\frac{R}{a}\Big)^2-C_{0,1}\Big(\frac{z}{b}\Big)^2\:\!\right],
\end{align}
where the coefficients $C_{\alpha, \beta}$ ($\alpha, \beta=0, 1$) for both the oblate and prolate cases are provided in Appendix~\ref{sec:A}. Note that $\Phi_{0, \mathrm{eff}}$ is equivalent to the potential of a 2D anisotropic harmonic oscillator. In the oblate case (i.e., $a>b$), the orbital angular frequency and the epicyclic frequencies for quasi-circular orbits inside the spheroid are derived from Eqs.~\eqref{eq:omegaphi2}, \eqref{eq:omegaR2}, and \eqref{eq:omegaz2} as 
\begin{align}
\omega_\phi^2/(2\pi \:\!G \rho_0)
&=C_{1,0}=
\frac{1-e^2}{e^2}\left(
\frac{\arcsin e}{e\sqrt{1-e^2}}-1
\right),
\label{eq:wphin0ob}
\\
\omega_{\mathrm{R}}^2/(2\pi \:\!G \rho_0)
&=4C_{1,0}=
4\frac{1-e^2}{e^2}\left(
\frac{\arcsin e}{e\sqrt{1-e^2}}-1
\right),
\label{eq:wRn0ob}
\\
\omega_{\mathrm{z}}^2/(2\pi \:\!G \rho_0)
&=\frac{C_{0,1}}{1-e^2}=
2e^{-2}\left(1-e^{-1}\sqrt{1-e^2}\arcsin e\right),
\label{eq:wzn0ob}
\end{align}
respectively, where $e=\sqrt{1-b^2/a^2}$. In the prolate case (i.e., $a<b$), the corresponding frequencies are given by
\begin{align}
\omega^2_{\phi}/(2\pi \:\!G \rho_0)
&=C_{1,0}=
e^{-2}
\left(
1-\frac{1-e^2}{e}\arctanh e
\right),
\label{eq:wphin0pro}
\\
\omega_{\mathrm{R}}^2/(2\pi \:\!G \rho_0)
&=4\:\!C_{1,0}=
4\:\!e^{-2}
\left(
1-\frac{1-e^2}{e}\arctanh e
\right),
\label{eq:wRn0pro}
\\
\omega_{\mathrm{z}}^2/(2\pi \:\!G \rho_0)
&=(1-e^2)\:\!C_{0,1}=
2\frac{1-e^2}{e^2}\left(-1+e^{-1}\arctanh e\right),
\label{eq:wzn0pro}
\end{align}
where $e=\sqrt{1-a^2/b^2}$. Equations~\eqref{eq:wphin0ob}--\eqref{eq:wzn0pro} show that these frequencies are independent of $R$. Figure~\ref{fig:n0} illustrates the dependence of these frequencies on $e$ for quasi-circular orbits within the Maclaurin spheroids. We find that all the squared frequencies are non-negative for all permissible values of $e$ in both the oblate and prolate cases. This indicates that epicyclic oscillations can occur near the equatorial plane. For the oblate case, both $\omega_{\phi}^2$ and $\omega_{\mathrm{R}}^2$ decrease monotonically with $e$, whereas $\omega_{\mathrm{z}}^2$ increases monotonically with $e$. Furthermore, we find that $\omega_{\phi}^2 \le \omega_{\mathrm{z}}^2 \le \omega_{\mathrm{R}}^2$ for $0\le e \le 0.955979$, while $\omega_{\phi}^2\le \omega_{\mathrm{R}}^2 \le \omega_{\mathrm{z}}^2 $ for $0.955979\le e \le 1$. For the prolate case, $\omega_{\phi}^2$ and $\omega_{\mathrm{R}}^2$ increase monotonically with $e$, whereas $\omega_{\mathrm{z}}^2$ decreases monotonically with $e$. Furthermore, we find that $\omega_{\mathrm{z}}^2\le \omega_{\phi}^2< \omega_{\mathrm{R}}^2$, with equality holding only for $e=0$.

For both prolate and oblate spheroids, Eqs.~\eqref{eq:wphin0ob}, \eqref{eq:wRn0ob}, \eqref{eq:wphin0pro}, and \eqref{eq:wRn0pro} imply that $\omega_\phi/\omega_{\mathrm{R}}=1/2$, which leads to a result, independent of $e$ and $R$, 
\begin{align}
\Delta \phi_{\mathrm{p}}=-\pi.
\end{align}
This result directly follows from the fact that the potential $\Phi_{0}(R,0)$ assumes a harmonic oscillator form (i.e., it is proportional to $R^2$ up to an additive constant). In other words, the result is derived under the conditions $\rho=\bar{\rho}=\rho_0$ (corresponding to a homogeneous regime) and $f=0$ in Eq.~\eqref{eq:fdef}. It is worth noting, however, that in the vacuum region outside the Maclaurin spheroid, $\Delta \phi_{\mathrm{p}}$ is generally positive~\citep{Kluzniak:2013,Bollimpalli:2022xgw}. Therefore, we can interpret the constant negativity of $\Delta \phi_{\mathrm{p}}$ in the internal region as a consequence of the extended mass effect in the homogeneous regime, without any additional influence from deformation.

\subsection{$n=1$}
\label{sec:5-1}
Next, we consider the case $n=1$. Then, the effective potential~\eqref{eq:Phieff} reduces to 
\begin{align}
\Phi_{1, \mathrm{eff}}={}
&\frac{l^2}{2R^2}-\frac{\pi \:\!G \rho_0 a^2}{2}\bigg[\:\!
C_{0,0}-2C_{1,0}\Big(\frac{R}{a}\Big)^2-2C_{0,1}\Big(\frac{z}{b}\Big)^2
\cr
&
+C_{2,0}\Big(\frac{R}{a}\Big)^4
+2C_{1,1}\Big(\frac{R}{a}\Big)^2\Big(\frac{z}{b}\Big)^2
+C_{0,2}\Big(\frac{z}{b}\Big)^4\:\!
\bigg],
\label{eq:Phi1eff}
\end{align}
where $C_{\alpha, \beta}$ for $\alpha, \beta=0, 1, 2$ are provided 
in Appendix~\ref{sec:A}. For $n=1$, the squares of the orbital angular frequency and the epicyclic frequencies, as given by Eqs.~\eqref{eq:omegaphi2}, \eqref{eq:omegaR2}, and \eqref{eq:omegaz2}, become 
\begin{align}
\omega_{\phi}^2/(2\pi \:\!G\rho_0)&=C_{1,0}-C_{2,0} \:\!\xi^2,
\label{eq:ophi1}
\\
\omega_{\mathrm{R}}^2/(2\pi \:\!G \rho_0)&=4 C_{1,0}-6 C_{2,0}\:\!\xi^2,
\label{eq:oR1}
\\
\omega_{\mathrm{z}}^2/(2\pi \:\!G \rho_0)&=C_{0,1}-C_{1,1}\:\!\xi^2,
\label{eq:oz1}
\end{align}
\begin{figure*}
\centering
\includegraphics[width=16cm,clip
]{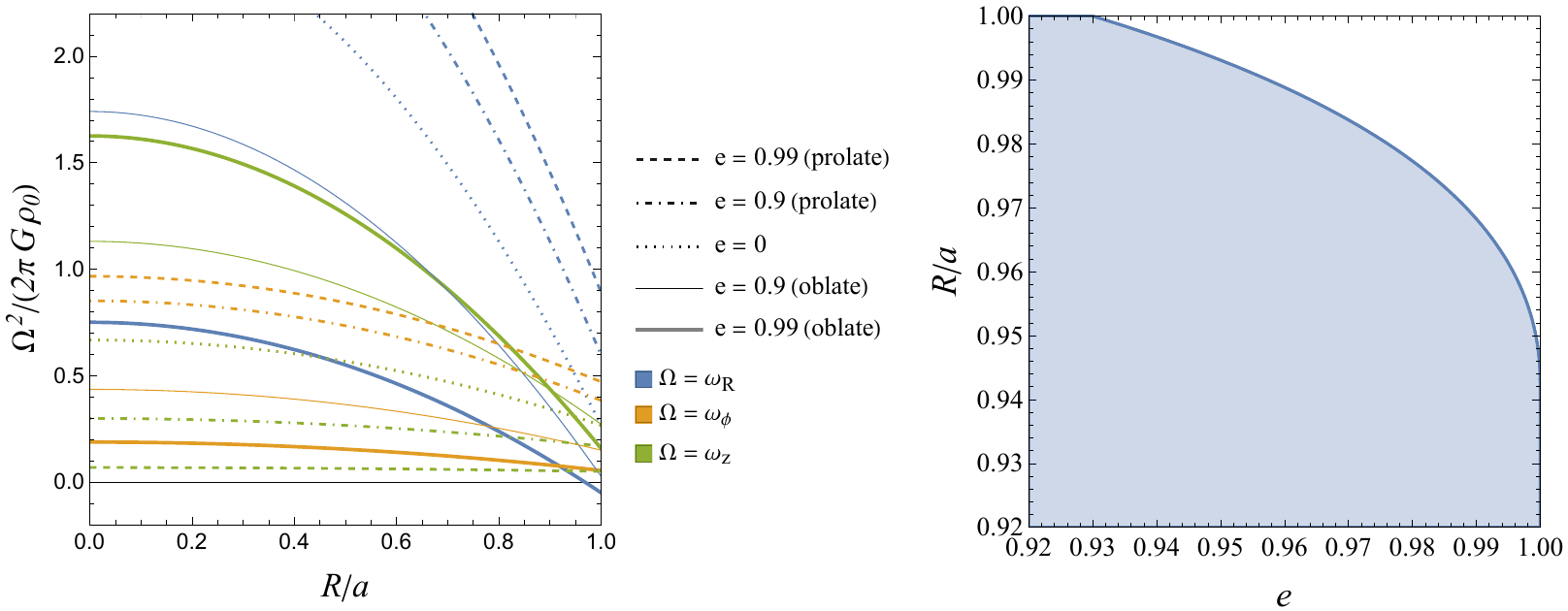}
\caption{Left panel: Plots of squared radial epicyclic frequencies (blue curves), squared orbital angular frequencies (orange curves), and squared vertical epicyclic frequencies (green curves) for $n=1$. Note that $\omega_{\mathrm{\phi}}=\omega_{\mathrm{z}}$ for the case $e=0$. Right panel: Parameter region inside the oblate spheroid where quasi-circular orbits exist (blue-shaded) and where they do not (unshaded). At the boundary of these two regions, $R=R_{\mathrm{ms}}$, marginal stable orbits appear. }
\label{fig:omegan1}
\end{figure*}
where $\xi=R/a$ as defined in Eq.~\eqref{eq:xi}. Unlike the $n=0$ case, these frequencies are no longer constants but functions of $\xi$. Figure~\ref{fig:omegan1} shows the dependence of the three squared frequencies on $R$. The values of $\omega_{\phi}^2$ and $\omega_{\mathrm{z}}^2$ remain positive for all $R$ in both the oblate and prolate cases. In contrast, $\omega_{\mathrm{R}}^2$ is positive for all $R$ in the prolate case but may become non-positive for the oblate case in the range $R\ge R_{\mathrm{ms}}$, where 
\begin{align}
R_{\mathrm{ms}}
=a\sqrt{\frac{2C_{1,0}}{3C_{2,0}}}
=a\left[\:\!
\frac{8 (e\sqrt{1-e^2}-\arcsin e)}{3\:\![\:\!
e\sqrt{1-e^2}(3+2e^2)-3\arcsin e
\:\!]}\:\!\right]^{1/2}.
\end{align}
If $R_{\mathrm{ms}}\le a$, then $R_{\mathrm{ms}}$ corresponds to the radius of a marginally stable circular orbit (MSCO) inside the oblate spheroid. The right panel in Fig.~\ref{fig:omegan1} shows that for the oblateness in the range $e_0\le e<1$ (with $e_0=0.929955685$ defined by $R_{\mathrm{ms}}|_{e=e_0}=a$), the condition $R_{\mathrm{ms}}<a$ holds. This implies that for $0<e<e_0$, the squared radial frequency $\omega_{\mathrm{R}}^2$ is positive for all $R\le a$, ensuring stable circular orbits throughout the spheroid. However, for $e_0\le e<1$, $\omega_{\mathrm{R}}^2$ remains positive only for $R<R_{\mathrm{ms}}$, becomes zero at $R= R_{\mathrm{ms}}$, and turns negative for $R_{\mathrm{ms}}<R\le a$.

The periapsis shift angle $\Delta \phi_{\mathrm{p}}$ is obtained by substituting Eqs.~\eqref{eq:ophi1} and \eqref{eq:oR1} into Eq.~\eqref{eq:Deltaphip} as follows:
\begin{align}
\Delta \phi_{\mathrm{p}}=2\pi \left(
\sqrt{\frac{C_{1,0}-C_{2,0} \xi^2}{4C_{1,0}-6 C_{2,0} \xi^2}}-1
\right).
\label{eq:dppn1}
\end{align}
Expanding the periapsis shift $\Delta \phi_{\mathrm{p}}$ in powers of $\xi$, we obtain
\begin{align}
\Delta \phi_{\mathrm{p}} = -\pi + O\left(\xi^2\right).
\end{align}
This leading term is the constant $-\pi$ (independent of $e$) because the first non-constant term in the expansion of the potential $\Phi_1$ in powers of $\xi$ is quadratic in $\xi$ (i.e., that of a harmonic oscillator).

By rewriting Eq.~\eqref{eq:dppn1} to include the parameter $f$, which accounts for the effects of deformation and its associated density inhomogeneities, we obtain
\begin{align}
\Delta \phi_{\mathrm{p}} = 2\pi \Biggl(\sqrt{1+\frac{3\:\!(1-\xi^2)}{1-\frac{3}{5}\:\!\xi^2}+f} -1 \Biggr),
\end{align}
where the local mass density is given by $\rho(\xi^2)=\rho_0(1-\xi^2)$ and the average mass density by $\bar{\rho}(\xi^2)=\rho_0(1-\tfrac{3}{5}\xi^2)$. The parameter $f$ is given as
\begin{align}
f=\mp\frac{12 \xi^2}{7\left(5-3\xi^2\right)^2}e^2+O(e^4),
\end{align}
with the upper sign for oblate deformations and the lower sign for prolate deformations. The leading-order term in the small-$e$ expansion is proportional to $e^2$, and therefore, $f=0$ for the spherically symmetric case (i.e., $e=0$). For oblate deformations, $f < 0$ at leading order, indicating that the oblateness suppresses the extended mass effect associated with the ratio of local density to average density. In contrast, for prolate deformations, $f > 0$ at leading order, meaning that the prolateness amplifies the extended mass effect arising from the mass density. 

\begin{figure*}
\centering
\includegraphics[width=15cm,clip]{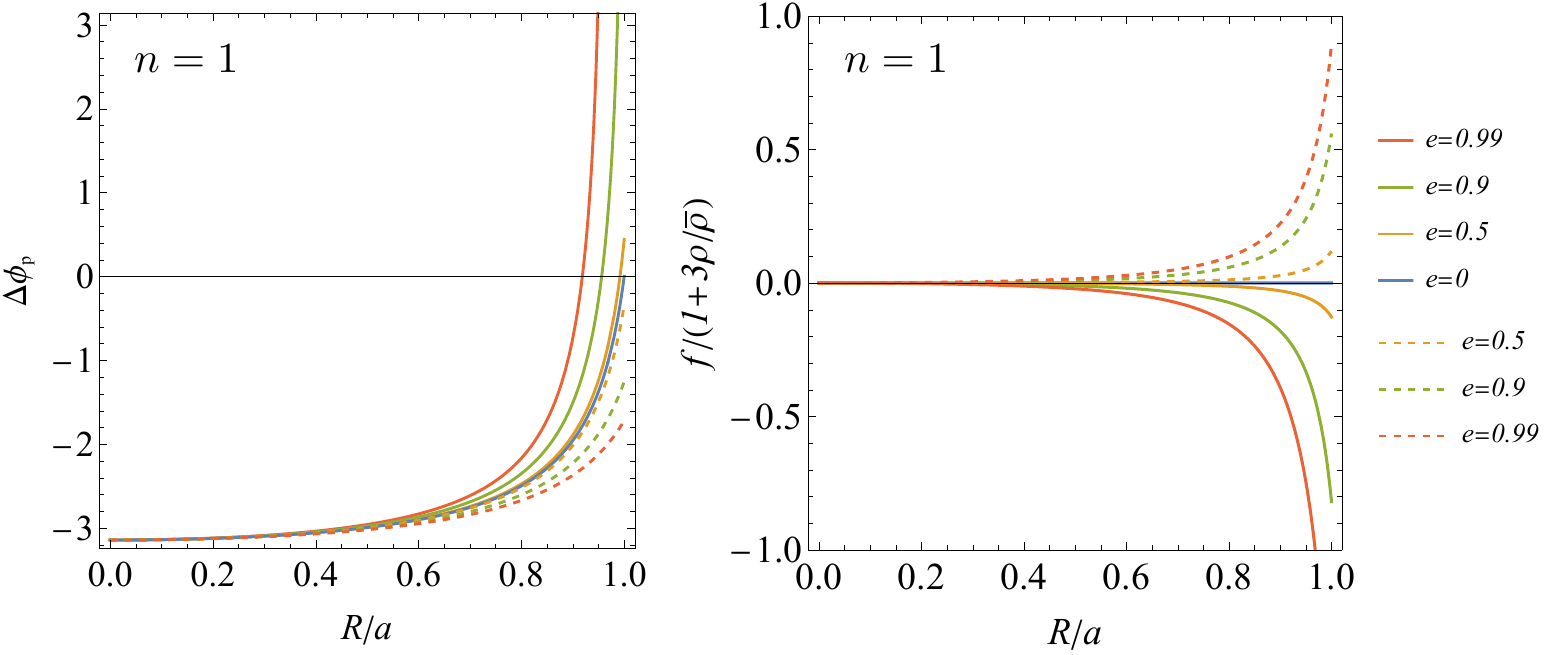}
\caption{Left panel: Periapsis shift angle $\Delta \phi_{\mathrm{p}}$ as a function of $R$ for $n=1$. Right panel: Values of the parameter $f$ as a function of $R$. Solid curves represent the oblate cases, whereas dashed curves represent the prolate cases.}
\label{fig:shiftn1}
\end{figure*}

Figure~\ref{fig:shiftn1} shows the $R$-dependence of $\Delta \phi_{\mathrm{p}}$ (left panel) and of the normalized parameter $f/(1+3\rho/\bar{\rho})$ (right panel) in the non-linear regime with respect to $e^2$. The left panel shows that, in both the oblate (solid curves) and prolate (dashed curves) cases, $\Delta \phi_{\mathrm{p}}$ increases monotonically with $R$, consistent with the result of $\partial(\Delta \phi_{\mathrm{p}})/\partial R \ge 0$ for all defined values of $e$. In particular, for $e=0$ (i.e., $f=0$), the effect of density inhomogeneity on $\Delta \phi_{\mathrm{p}}$ evidently appears to increase from $-\pi$ toward $0$, which can be interpreted as a reduction in the extended mass effect resulting from a decrease in the ratio of local to average density. For $e \neq 0$, a similar monotonic increase can be attributed to the same factors; furthermore, the additional influence of deformation arises from density inhomogeneity, an effect that is absent in the homogeneous case. For the prolate case, an increasing $e$ shifts $\Delta \phi_{\mathrm{p}}$ in the negative direction, indicating that the prolateness amplifies the extended mass effect arising from the mass density. In contrast, for the oblate case, an increasing $e$ shifts $\Delta \phi_{\mathrm{p}}$ in the positive direction, such that near the surface $\Delta \phi_{\mathrm{p}}$ becomes positive, meaning that the oblateness suppresses the extended mass effect associated with the mass density. Moreover, for $e\ge e_0$, as $R$ approaches $R_{\mathrm{ms}}$, $\Delta \phi_{\mathrm{p}}$ diverges to infinity because $\omega_{\mathrm{R}}$ simultaneously tends to zero. The right panel quantitatively demonstrates that the deformation effect becomes increasingly significant as $R$ approaches the surface. These findings agree with the behaviour of the periapsis shift and indicate that the effects of deformation and the resulting density inhomogeneity increase monotonically from the center to the surface of the spheroid.

\subsection{$n=2$}
\label{sec:5-2}
\begin{figure*}
\centering
\includegraphics[width=16cm,clip]{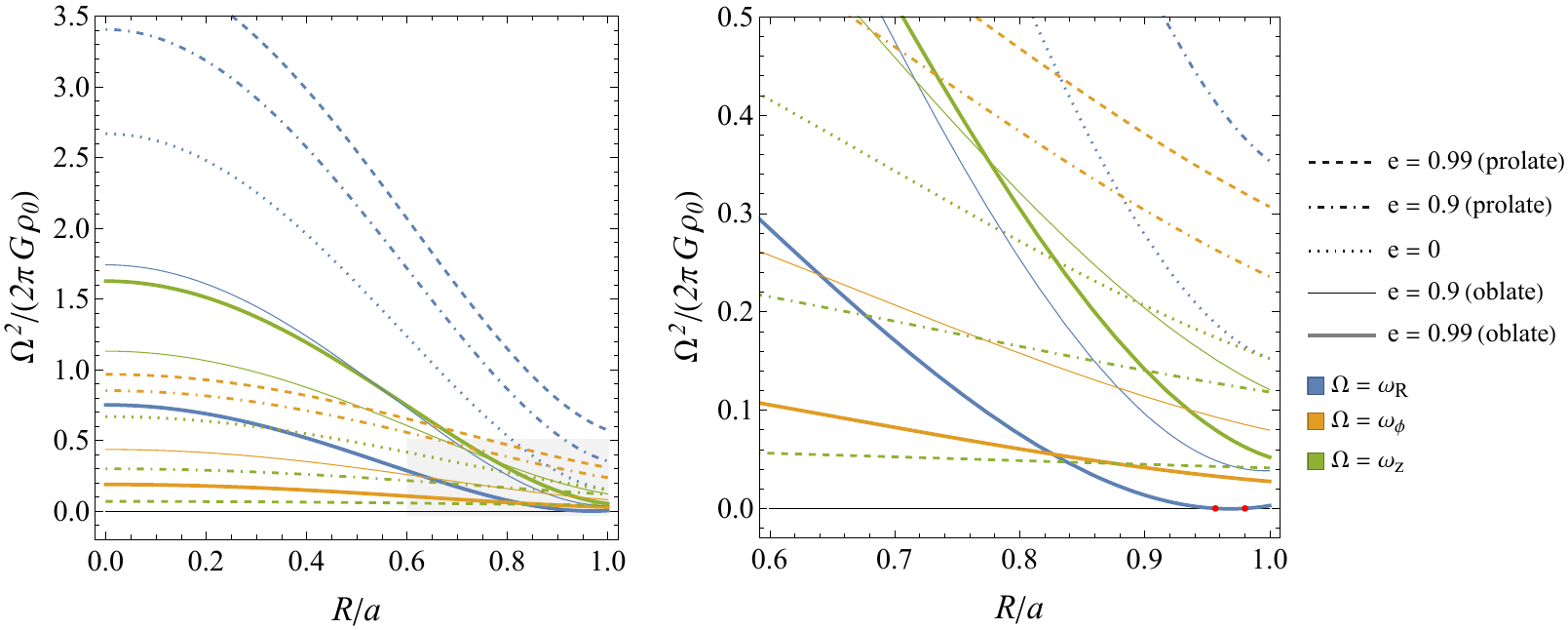}
\caption{Plots of the squared orbital angular and epicyclic frequencies for $n=2$. The right panel shows a magnified view of the shaded region shown in the left panel; the two red dots mark the radii where $\omega_{\mathrm{R}}^{2}=0$ for the oblate sequence with $e=0.99$, corresponding to the MSCOs.}
\label{fig:omegan2}
\end{figure*}

We now consider the case $n=2$. In this case, the effective potential~\eqref{eq:Phieff} reduces to 
\begin{align}
\Phi_{2, \mathrm{eff}}={}&\frac{l^2}{2R^2}
-\frac{\pi \:\!G \rho_0 a^2}{3}\bigg[\:\!
C_{0,0}-3C_{1,0}\Big(\frac{R}{a}\Big)^2-3C_{0,1}\Big(\frac{z}{b}\Big)^2
\nonumber
\\
&
+3C_{2,0}\Big(\frac{R}{a}\Big)^4+6C_{1,1}\Big(\frac{R}{a}\Big)^2\Big(\frac{z}{b}\Big)^2+3C_{0,2}\Big(\frac{z}{b}\Big)^4
\nonumber
\\
&-C_{3,0}\Big(\frac{R}{a}\Big)^6-3C_{2,1}\Big(\frac{R}{a}\Big)^4\Big(\frac{z}{b}\Big)^2
\nonumber
\\
&
-3C_{1,2}\Big(\frac{R}{a}\Big)^2\Big(\frac{z}{b}\Big)^4-C_{0,3}\Big(\frac{z}{b}\Big)^6\:\!
\bigg],
\end{align}
where the coefficients $C_{\alpha, \beta}$ for $\alpha, \beta=0, 1, 2, 3$ are given in Appendix~\ref{sec:A}. The squared orbital angular and epicyclic frequencies given by Eqs.~\eqref{eq:omegaphi2}, \eqref{eq:omegaR2}, and \eqref{eq:omegaz2} are expressed as follows:
\begin{align}
\omega_{\phi}^2/(2\pi \:\!G\rho_0)
&=
C_{1,0}-2C_{2,0}\:\!\xi^2+C_{3,0}\:\!\xi^4,
\\
\omega_{\mathrm{R}}^2/(2\pi \:\!G \rho_0)
&=
4C_{1,0}-12C_{2,0}\:\!\xi^2+8C_{3,0}\:\!\xi^4,
\\
\omega_{\mathrm{z}}^2/(2\pi \:\!G \rho_0)
&=
C_{0,1}-2C_{1,1}\:\!\xi^2+C_{2,1}\:\!\xi^4.
\end{align}
Figure~\ref{fig:omegan2} shows the plots of these squared frequencies. Notably, in the oblate case at $e=0.99$, $\omega^2_{\mathrm{R}}$ exhibits two zero crossings, implying the existence of two distinct MSCOs. 

The periapsis shift angle is given by
\begin{align}
\Delta \phi_{\mathrm{p}}=2\pi \left(
\sqrt{\frac{C_{1,0}-2C_{2,0}\xi^2+C_{3,0}\xi^4}{4C_{1,0}-12C_{2,0}\xi^2+8C_{3,0}\xi^4}}-1
\right).
\end{align}
Expanding $\Delta \phi_{\mathrm{p}}$ in powers of $\xi$, we obtain $\Delta \phi_{\mathrm{p}} = -\pi + O\left(\xi^2\right)$, since the first non-zero correction in the expansion of $\Phi_2$ is of order $\xi^2$. This expression can be rewritten to include the parameter $f$ as 
\begin{align}
\Delta \phi_{\mathrm{p}}=2\pi \left(
\sqrt{1+\frac{3\:\!(1-\xi^2)^2}{1-\frac{6}{5} \xi^2+\frac{3}{7}\xi^4}+f}-1
\right),
\end{align}
where the local mass density is given as $\rho(\xi^2)=(1-\xi^2)^2$ and the averaged mass density as $\bar{\rho}(\xi^2)=\rho_0\left(1-\frac{6}{5}\xi^2+\frac{3}{7}\xi^4\right)$. The parameter $f$ is given by
\begin{align}
f=\mp \frac{8 \xi^2 \left(63-70 \xi^2+15 \xi^4\right)}{3\left(35-42 \xi^2+15 \xi^4\right)^2} \:\!e^2+O(e^4),
\end{align}
with the upper sign for oblate deformations and the lower sign for prolate deformations. For oblate deformations, the leading term in the $e^2$ expansion results in $f<0$, thereby suppressing the extended mass effect arising from the mass density; conversely, for prolate deformations, $f>0$, thereby amplifying the extended mass effect.

\begin{figure*}
\centering
\includegraphics[width=15cm,clip]{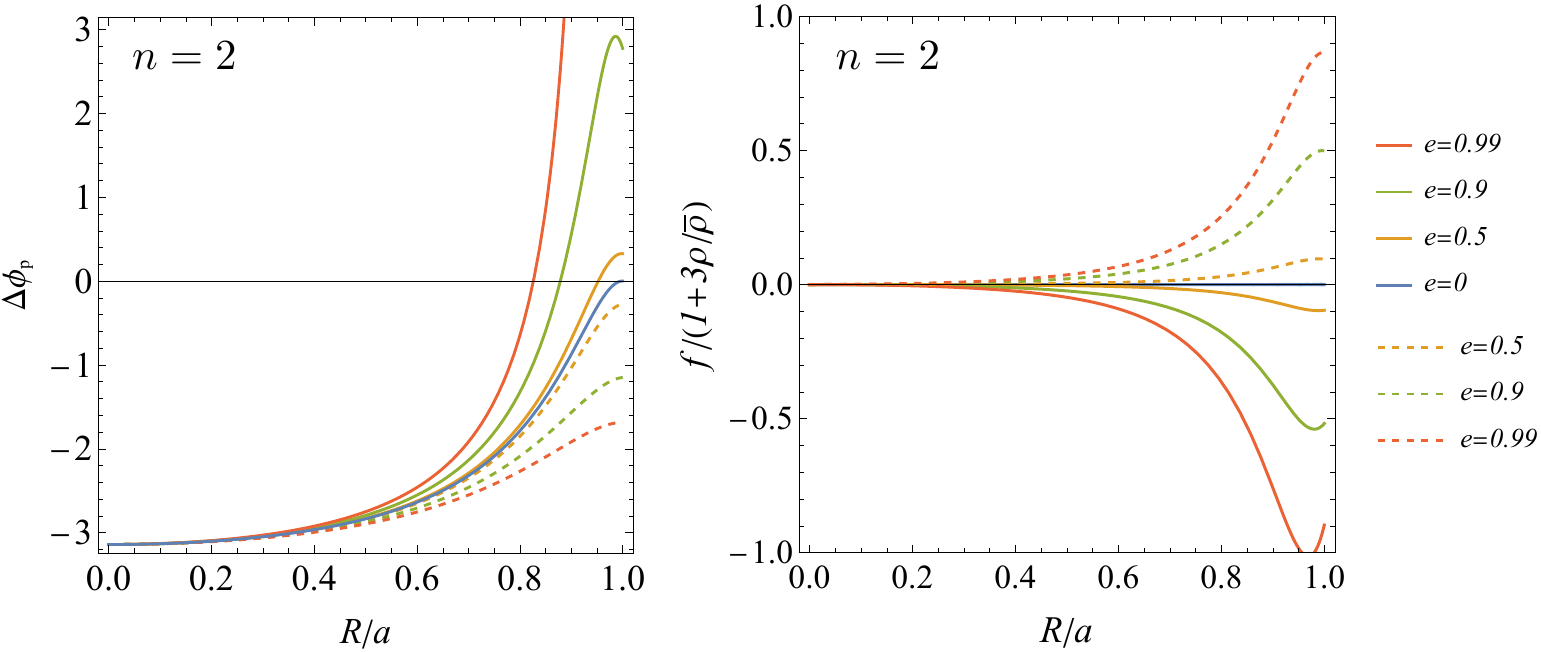}
\caption{Left panel: Periapsis shift angle $\Delta \phi_{\mathrm{p}}$ as a function of $R$ for $n=2$. Right panel: Values of the parameter $f$ as a function of $R$. Solid curves represent the oblate cases, while dashed curves represent the prolate cases.}
\label{fig:shiftn2}
\end{figure*}

Figure~\ref{fig:shiftn2} shows the $R$-dependence of $\Delta \phi_{\mathrm{p}}$ in the left panel and of the normalized parameter $f/(1+3\rho/\bar{\rho})$ in the right panel. In the prolate case (dashed curves), $\Delta \phi_{\mathrm{p}}$ increases monotonically with $R$ and changes in the negative direction as $e$ increases. Furthermore, the normalized parameter indicates that the deformation effect becomes more pronounced as $R$ increases. In contrast, in the oblate case (solid curves), $\Delta \phi_{\mathrm{p}}$ consistently changes in the positive direction as $e$ increases, but its $R$-dependence varies qualitatively with $e$. The normalized parameter further suggests that the deformation effect suppresses the extended mass effect.

\begin{figure*}
\centering
\includegraphics[width=15cm,clip]{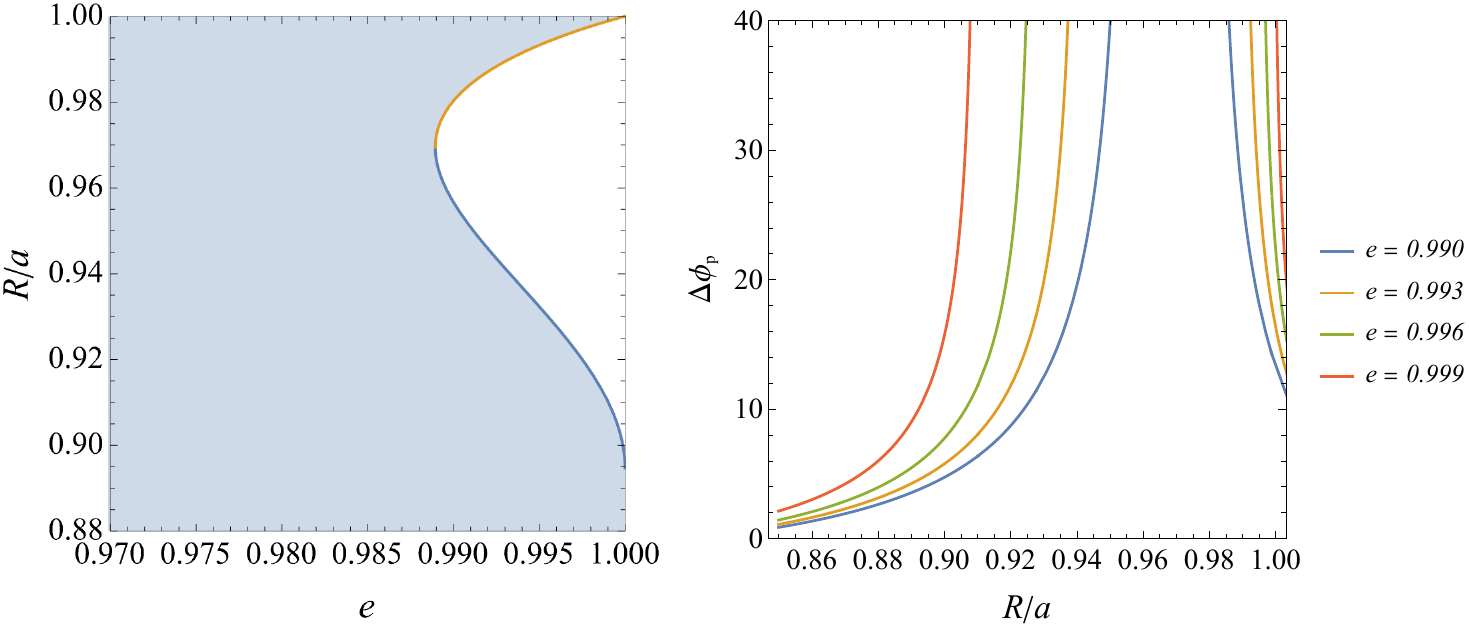}
\caption{
Left panel: Region in the $R$--$e$ plane for $n=2$ in the oblate case, showing where quasi-circular orbits exist (blue-shaded) and where they do not (unshaded). The boundary, depicted by the blue and orange curves, corresponds to the MSCOs, with their intersection marking the critical value $e = 0.988937$. Right panel: Plots of $\Delta \phi_{\mathrm{p}}$ for several values of $e$ (with $e> 0.988937$), each curve exhibiting divergence at two distinct radii.
}
\label{fig:oscon2}
\end{figure*}

The behaviour of $\Delta \phi_{\mathrm{p}}$ undergoes a significant transition at the critical value $e_0=0.988937349$. The left panel of Fig.~\ref{fig:oscon2} shows the region in the $R$--$e$ plane where quasi-circular orbits exist (blue-shaded) and where they do not (unshaded). When $e<e_0$, quasi-circular orbits exist over the entire range of $R$, whereas for $e_0 \leq e < 1$, a region in $R$ emerges where quasi-circular orbits are absent (i.e., where $\omega^2_{\mathrm{R}}<0$). Note that, unlike in the $n=1$ case where the MSCO appears at the spheroid's surface, at $e = e_0$, it occurs at $R/a = 0.969442391$.
Consequently, for $e$ in the range $[e_0,1)$, $\Delta \phi_{\mathrm{p}}$ diverges at those radii corresponding to the two MSCOs (see the right panel of Fig.~\ref{fig:oscon2}).

\begin{figure}
\centering
\includegraphics[width=\columnwidth]{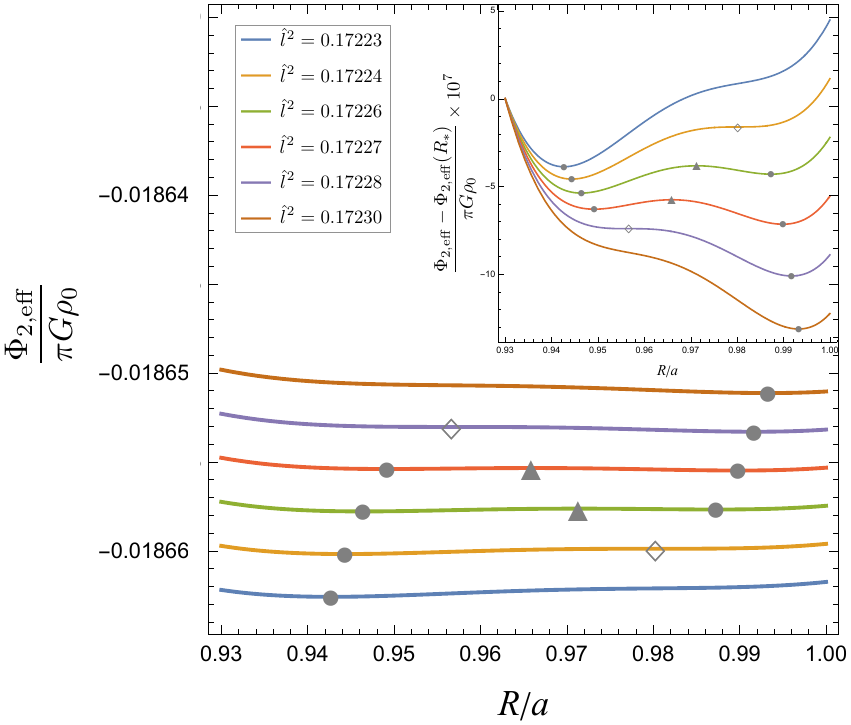}
\caption{Main panel: Equatorial effective potential $\Phi_{2, \mathrm{eff}}(R,0)$ for an inhomogeneous spheroid with $n = 2$ and oblateness $e = 0.99$, plotted for six values of $\hat{l}^2\equiv l^2/(\pi G \rho_0 a^4)$. Inset: The same curves after subtracting, curve-by-curve, the fixed
reference value $\Phi_{2,\mathrm{eff}}(R_{\ast},0)$ with $R_*=0.93\,a$. Filled circles mark the locally stable minima, filled triangles indicate the locally unstable maxima, and open diamonds mark the radii of the inner and outer MSCOs.}
\label{fig:potentials}
\end{figure}

Figure~\ref{fig:potentials} shows $\Phi_{2,\mathrm{eff}}(R,0)$ for oblateness $e = 0.99$. Two filled‐circle sequences identify distinct families of stable circular orbits: the left-hand sequence forms the inner branch and the right-hand sequence the outer branch. As $l^{2}$ increases, the guiding radius on the inner branch increases monotonically until the branch terminates at the inner MSCO, marked by the left-hand open diamond. Once slightly displaced from this MSCO, the particle eventually crosses the spheroidal surface. Conversely, as $l^{2}$ decreases, the guiding radius on the outer branch decreases monotonically until the branch terminates at the outer MSCO, marked by the right-hand open diamond. A slight displacement from the outer MSCO drives the particle inward, confining it between the outer MSCO and the inner centrifugal barrier.

The radial interval between the two MSCOs contains only a sequence of
unstable circular orbits (filled triangles), for which $\omega_{\mathrm{R}}^{2}<0$; hence no epicyclic oscillations can occur there. A particle placed on one of these unstable circular orbits and given a small radial perturbation inevitably departs: if its specific energy
exceeds the surface value $\Phi_{2,\mathrm{eff}}(R=a,0)$ it escapes to the exterior, whereas for lower energy if it is reflected by the central or surface barrier and remains trapped between them.

In the oblate case, we focus on $e=0.9$ as shown in Fig.~\ref{fig:shiftn2}. Here, $\Delta \phi_{\mathrm{p}}$ exhibits a peak and then decreases near the surface. This behaviour is partly because the extended mass effect weakens in the surface region due to the low local density. Similarly, the normalized function $f/(1+3\rho/\bar{\rho})$ with negative values also exhibits a minimum near the surface before increasing with $R$, indicating a reduced contribution from the deformation effect. The combined effects of these factors determine the behaviour of the shift angle in the near-surface region.

\subsection{$n=3$}
\label{sec:5-3}
\begin{figure*}
\centering
\includegraphics[width=15cm,clip]{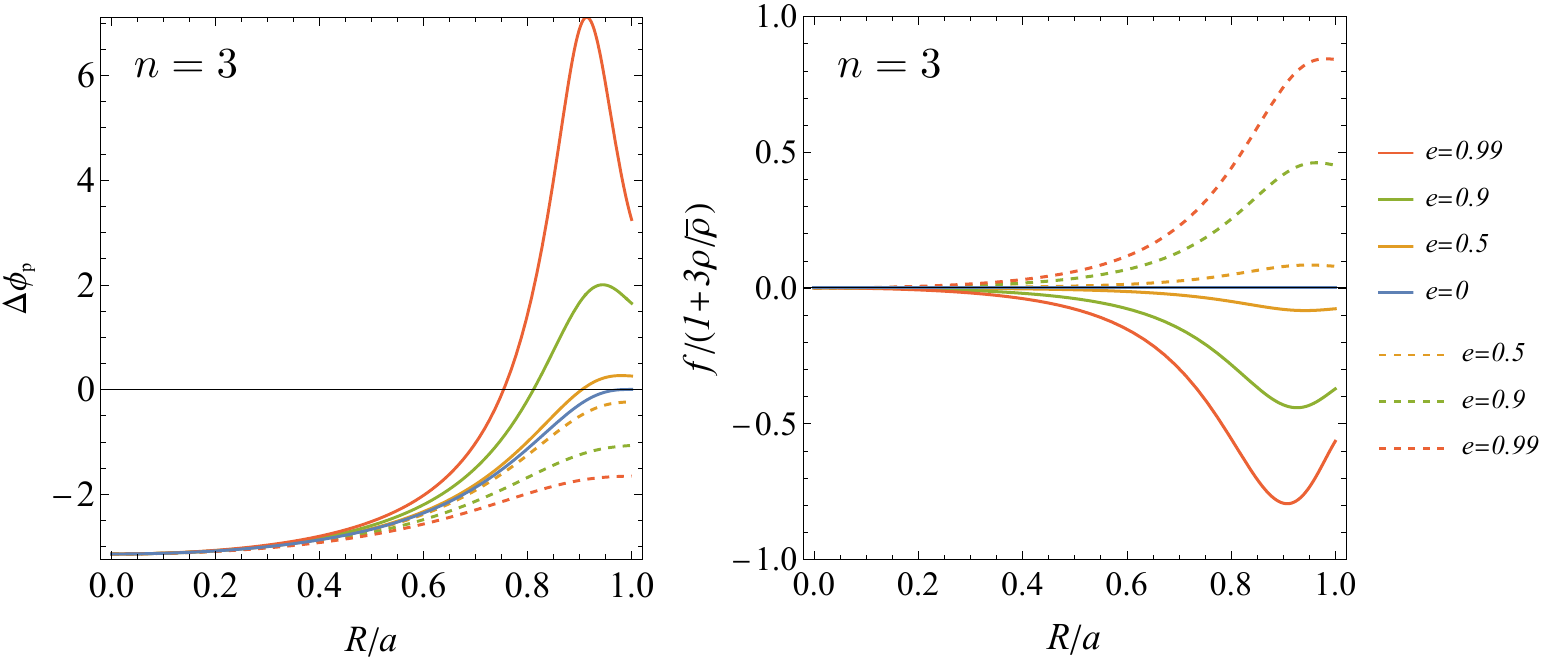}
\includegraphics[width=15cm,clip]{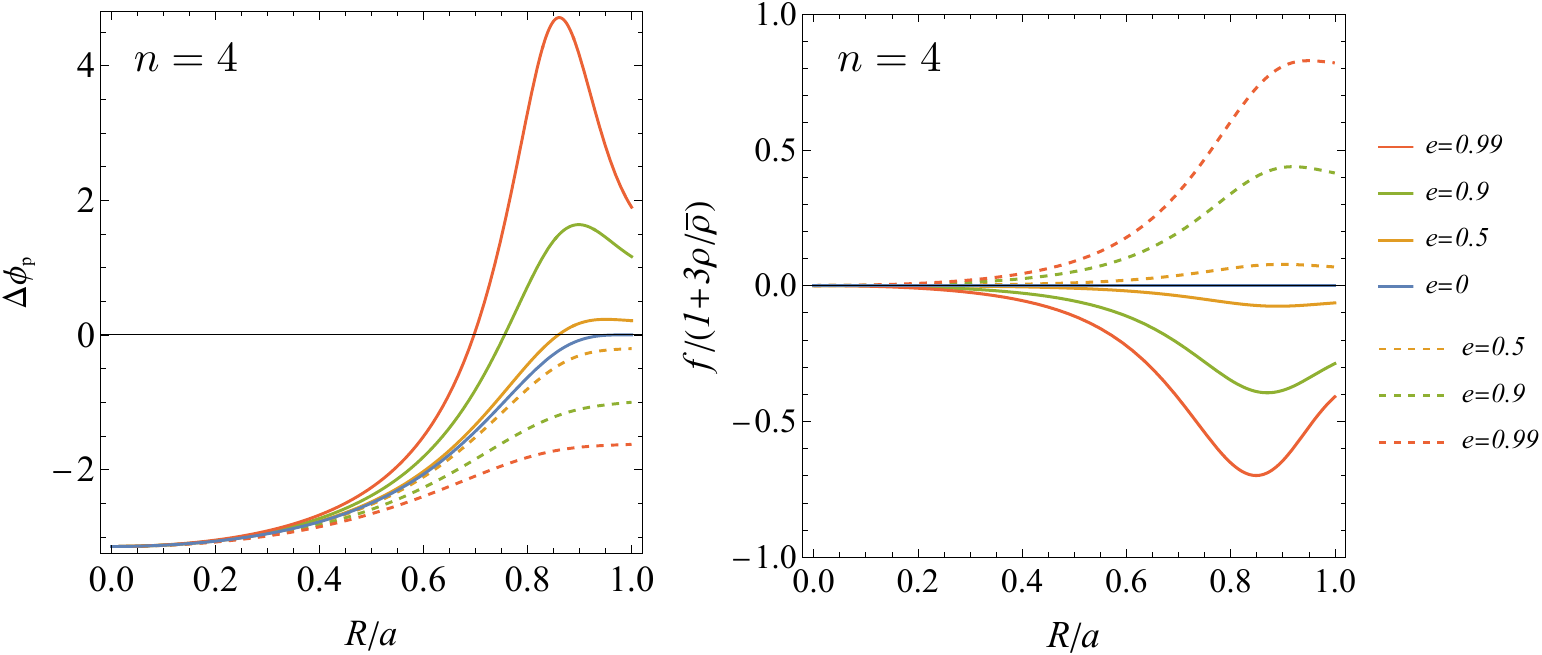}
\caption{Upper left panel: $\Delta \phi_{\mathrm{p}}$ for $n=3$. Upper right panel: $f/(1+3\rho/\bar{\rho})$ for $n=3$. Lower left panel: $\Delta \phi_{\mathrm{p}}$ for $n=4$. Lower right panel: $f/(1+3\rho/\bar{\rho})$ for $n=4$. Solid curves represent the oblate cases, while dashed curves represent the prolate cases.}
\label{fig:shiftn34}
\end{figure*}

In the case $n=3$, the effective potential $\Phi_{3, \mathrm{eff}}$ yields the following expressions for the orbital angular and epicyclic frequencies:
\begin{align}
\omega_{\phi}^2/(2\pi \:\!G\rho_0)
&=
C_{1,0}-3C_{2,0}\:\!\xi^2+3C_{3,0}\:\!\xi^4-C_{4,0}\:\!\xi^6,
\\
\omega_{\mathrm{R}}^2/(2\pi \:\!G \rho_0)
&=4C_{1,0}-18C_{2,0}\:\!\xi^2+24C_{3,0}\:\!\xi^4-10C_{4,0}\:\!\xi^6,
\\
\omega_{\mathrm{z}}^2/(2\pi \:\!G \rho_0)
&=C_{0,1}-3C_{1,1}\:\!\xi^2+3C_{2,1}\:\!\xi^4-C_{3,1}\:\!\xi^6. 
\end{align}
The upper panels of Fig.~\ref{fig:shiftn34} show $\Delta \phi_{\mathrm{p}}$ and $f/(1+3\rho/\bar{\rho})$.
Note, however, that MSCOs do not appear, and quasi-circular orbits exist throughout the entire range $R\le a$. 

\subsection{$n=4$}
\label{sec:5-4}
In the case $n=4$, the effective potential $\Phi_{4, \mathrm{eff}}$ yields the following expressions for the orbital angular and epicyclic frequencies:
\begin{align}
\omega_{\phi}^2/(2\pi \:\!G\rho_0)
={}&C_{1,0}-4C_{2,0}\:\!\xi^2+6C_{3,0}\:\!\xi^4-4C_{4,0}\:\!\xi^6+C_{5,0}\:\!\xi^8,
\\
\omega_{\mathrm{R}}^2/(2\pi \:\!G \rho_0)
={}&
4C_{1,0}-24C_{2,0}\:\!\xi^2+48C_{3,0}\:\!\xi^4
\nonumber
\\
&
-40C_{4,0}\:\!\xi^6+12 C_{5,0}\:\!\xi^8,
\\
\omega_{\mathrm{z}}^2/(2\pi \:\!G \rho_0)
={}&
C_{0,1}-4C_{1,1}\:\!\xi^2+6C_{2,1}\:\!\xi^4-4C_{3,1}\:\!\xi^6+C_{4,1}\:\!\xi^8.
\end{align}
The lower panels of Fig.~\ref{fig:shiftn34} show $\Delta \phi_{\mathrm{p}}$ and $f/(1+3\rho/\bar{\rho})$. No MSCO appears, and quasi-circular orbits exist throughout the entire range $R\le a$.

\section{Summary and discussion}
\label{sec:6}
Within the framework of Newtonian theory, we have investigated the effects of both extended mass and gravitational source deformation on the periapsis shift phenomenon in quasi-circular orbits. Specifically, we focused on an ellipsoidal source consisting of uniform-density homeoid layers. By assuming a specific density profile and a spheroidal configuration, we analysed the quasi-circular orbits confined to the reflection-symmetric plane of the interior gravitational potential. Using this approach, we derived expressions for the orbital angular frequency, the epicyclic frequencies, and the periapsis shift angle. Furthermore, by varying the parameters inherent in the density distribution, we provide a detailed analysis of how spheroidal deformation and associated density inhomogeneities affect these frequencies and the shift angle. 

We show that the effects of extended mass and source deformation on quasi-circular orbits vary significantly with the degree of density inhomogeneity and the extent of spheroidal deformation. In the case of uniform density, the internal gravitational potential reduces to that of a harmonic oscillator; consequently, both the orbital angular frequency and epicyclic frequencies become independent of the guiding orbit radius, and the periapsis shift angle remains constant at $-\pi$. This implies that for a uniform-density spheroid, the deformation does not affect the periapsis shift. In contrast, for inhomogeneous density distributions, both the orbital angular frequency and the epicyclic frequencies depend on the guiding orbit radius and the deformation parameter, causing the shift angle to deviate from the uniform-density value. As in the spherically symmetric system, the extended mass effect influences the shift angle through the ratio of the local density to the average density, with a higher local density contributing to a more negative shift. Moreover, we find that in the case of oblate deformation, this effect is suppressed, whereas in the case of prolate deformation, it is amplified.

We also analysed the stability of quasi-circular orbits in this system by examining the epicyclic frequencies. In the uniform-density case, the quasi-circular orbits remain stable regardless of the guiding orbit radius or the deformation parameter. In contrast, when the density is inhomogeneous---especially under oblate deformation---the radial epicyclic frequency becomes zero, indicating that MSCOs appear for $e_0 \le e < 1$. This phenomenon seems to occur only for $n=1$ and $n=2$. Specifically, for $n=1$, the critical value is $e_0 = 0.929955685$, which agrees with the oblateness at which MSCOs emerge in the external vacuum region~\citep{Bollimpalli:2022xgw}. This agreement occurs because, in both the interior region and the external vacuum region, the MSCO emerges at the spheroid's surface when $e=e_0$. In contrast, for $n=2$, the critical value is $e_0 = 0.988937349$, and no such correspondence with the external vacuum region's oblateness is observed. Notably, the MSCOs---often misinterpreted as a phenomenon unique to general relativity---emerge under Newtonian conditions due to extended mass effects and density inhomogeneity. Meanwhile, under prolate deformation, MSCOs are not formed.

To evaluate the astrophysical relevance of our results, we consider the effect of the dark matter distribution on Mercury's motion in the Solar system as a concrete example. The mean mass density inside Mercury's orbital semimajor axis $a_{\mathrm{Mercury}} = 5.79 \times 10^{12}\mathrm{\,cm}$~\citep{Williams:2025} is determined by the mass of the Sun and is approximately $\bar{\rho}_{\mathrm{Mercury}}=3M_{\odot}/(4\pi a_{\mathrm{Mercury}}^3)=2.5\times 10^{-6}\mathrm{\, g/cm^3}$. On the other hand, the upper limit on the local dark matter density is $\rho_{\mathrm{DM}}=9.3\times 10^{-18} \mathrm{\, g/cm^3}$~\citep{Pitjev:2013}. Thus, the ratio $\rho_{\mathrm{DM}}/\bar{\rho}_{\mathrm{Mercury}}$ is estimated to be $O(10^{-12})$, placing it in the diffuse regime. Then, the correction to the orbital frequency induced by dark matter is of the order $\omega_\phi^{(\mathrm{DM})}\sim 10^{-13} \mathrm{\,rad/s}$,%
\footnote{Assuming the scale of the dark matter distribution is $a=10$--$100 \mathrm{\,au}$, we have $R_{\mathrm{Mercury}}/a \ll 1$. In this regime, $C_{1,0}=2/3$, and the correction to the orbital frequency can be estimated as $\omega_\phi^{(\mathrm{DM})} \sim \sqrt{2\pi G \rho_0}$.}
which is negligible compared to the orbital frequency of Mercury, $\omega_\phi^{(\mathrm{Mercury})} = 8.3 \times 10^{-7}\ \mathrm{\, rad/s}$~\citep{Williams:2025}. The corresponding perihelion shift, $\Delta \phi_{\mathrm{p}}=-3.5\times 10^{-11}\mathrm{\,rad/orbit}$, is negligible compared to the observed value of the relativistic correction, $42.98 \mathrm{\,arcsec/cy}\approx 5.0\times 10^{-7}\mathrm{\,rad/orbit}$~\citep{Park:2017}. Therefore, the correction from dark matter is negligible compared to the observed perihelion shift of Mercury, and the deformation effect, as a higher-order contribution, is well below the current detectability limit in the Solar system.

As another astrophysical application, we consider the effect of dark matter on the orbit of the S2 star around the Galactic center. The S2 star has an orbital semimajor axis of $a_\mathrm{S2}= 970\mathrm{\,au} \approx 1.45 \times 10^{16} \mathrm{\, cm}$ and eccentricity $e_{\mathrm{S2}} = 0.88$, and orbits the supermassive object at the center of the Milky Way, with mass $M_\bullet = 4.3 \times 10^6 M_\odot$~\citep{GRAVITY:2020}. Although our quasi-circular approximation does not strictly apply to such a highly eccentric orbit, we can still estimate the order of magnitude of the effect by considering the mean enclosed density. The mean mass density enclosed within $a_\mathrm{S2}$, set by the central mass, is given by $\bar{\rho}_\mathrm{S2} = \frac{3 M_\bullet}{4\pi a_\mathrm{S2}^3}
    \approx 6.7 \times 10^{-10}\mathrm{\,g/cm^3}$. For illustration, if we assume that the maximum allowed amount of dark matter (corresponding to a mass fraction of 0.03\% or $1200M_\odot$~\citep{GRAVITY:2020}) is uniformly distributed within the S2 orbital radius, the resulting density is $\rho_{\mathrm{DM}} = 1.9 \times 10^{-13}\mathrm{\, g/cm^3}$. This yields a periapsis shift correction of $\Delta \phi_{\mathrm{p}} \approx -2.6 \times 10^{-3} \mathrm{\,rad/orbit}$, which is comparable to the general relativistic contribution ($3.5\times 10^{-3} \mathrm{\,rad/orbit}$). The correction due to the deformation effect is expected to be of a similar order, or possibly smaller, depending on the details of the deformation parameter. This implies that, if a significant amount of deformation is present in the mass distribution near the Galactic Center, its impact on the periapsis shift could in principle compete with the relativistic effect. More realistically, in standard scenarios, such as those based on the NFW halo profile (see e.g.~\citealt{Gnedin:2004}), the dark matter density at the Galactic center is expected to be $O(10^{-19})\mathrm{\,g/cm^3}$. However, if the formation of a dark matter spike due to adiabatic growth of the central black hole is considered, the density can reach much higher values, potentially up to $O(10^{-16})\mathrm{\,g/cm^3}$ or even greater~\citep{Gondolo:1999ef, Sadeghian:2013laa,Lacroix:2018zmg}, in which case the extended mass effect (i.e., the spherically symmetric component) and the associated deformation effect (the non-spherical correction), would become more pronounced due to the localization of the dark matter density.

We will report the results of our post-Newtonian analysis in a separate paper, which examines the competition between the general-relativistic effects and the combined extended mass and deformation effects on the periapsis shift of quasi-circular orbits in a system featuring a black hole at the centre of a spheroidal mass distribution. Furthermore, the extended mass and deformation effects on the periapsis shift of quasi-circular orbits in static, axisymmetric spacetimes warrant future work~\citep{Iorio:2023ajy}. In addition, generalizing the general formula for the periapsis shift of quasi-circular orbits in static, spherically symmetric systems~\eqref{eq:DphipGR} to static, axisymmetric spacetimes is a promising topic for future research.

\section*{Acknowledgements}
The author gratefully acknowledges the useful comments and discussions provided by Kota Ogasawara, Tomohiro Harada, Masashi Kimura, Keiju Murata, Takeshi Chiba, Masaya Amo, Akihito Katsumata, Takashi Mishima, Yohsuke Takamori, Takaaki Ishii, Hideki Ishihara, Ken-ichi Nakao, Yasufumit Kojima, Masaaki Takahashi, Rohta Takahashi, Shinji Koide, Hirotaka Yoshino, and So Aina. The author would like to thank the anonymous referee for their constructive comments, which improved the quality of this manuscript. This work was supported by JSPS KAKENHI Grant Nos.~JP22K03611, JP23KK0048, and JP24H00183.

\appendix

\section*{DATA AVAILABILITY}
No data were created or analysed in this article.

\section{Explicit form of $C_{\alpha, \beta}$}
\label{sec:A}
\subsection{Oblate case ($a>b$)}
We list the explicit forms of $C_{\alpha, 0}$ (for $\alpha=0, 1, 2, 3, 4, 5$) 
and $C_{\alpha, 1}$ (for $\alpha=0, 1, 2, 3, 4$) in the oblate case as follows:
\begin{align}
C_{0,0}={}&\frac{2\sqrt{1-e^2}}{e}\arcsin e,
\\
C_{1,0}={}&
\frac{1-e^2}{e^2}\left(\frac{\arcsin e}{e\sqrt{1-e^2}}-1\right),
\\
C_{2,0}={}&
\frac{3(1-e^2)}{4e^4}\left(
\frac{\arcsin e}{e\sqrt{1-e^2}}
-1-\frac{2e^2}{3}
\right),
\\
C_{3,0}={}&
\frac{15(1-e^2)}{24 e^6}\left(
\frac{\arcsin e}{e\sqrt{1-e^2}}
-1-\frac{2e^2}{3}-\frac{8e^4}{15}
\right),
\\
C_{4,0}={}&
\frac{35(1-e^2)}{64 e^8}\left(
\frac{\arcsin e}{e\sqrt{1-e^2}}-1-\frac{2e^2}{3}-\frac{8e^4}{15}-\frac{16e^6}{35}
\right),
\\
C_{5,0}={}&
\frac{63(1-e^2)}{128 e^{10}}
\bigg(
\frac{\arcsin e}{e\sqrt{1-e^2}}-1-\frac{2e^2}{3}-\frac{8e^4}{15}-\frac{16 e^6}{35}
\nonumber
\\
&\quad -\frac{128 e^8}{315}
\bigg).
\end{align}

The coefficients $C_{\alpha,1}$ are given by
\begin{align}
C_{0,1}={}&
\frac{2(1-e^2)}{e^2}\left(1-\frac{\sqrt{1-e^2}}{e}\arcsin e
\right),
\\
C_{1,1}={}&
\frac{3(1-e^2)}{e^4}\left(
1-\frac{e^2}{3}-\frac{\sqrt{1-e^2}}{e} \arcsin e
\right),
\\
C_{2,1}={}&
\frac{15(1-e^2)}{4e^6}\left(
1-\frac{e^2}{3}-\frac{2e^4}{15}-\frac{\sqrt{1-e^2}}{e}\arcsin e
\right),
\\
C_{3,1}={}&
\frac{35(1-e^2)}{8e^8}\bigg(
1-\frac{e^2}{3}-\frac{2 e^4}{15}-\frac{8 e^6}{105}
\nonumber
\\
&\quad
-\frac{\sqrt{1-e^2}}{e} \arcsin e
\bigg),
\\
C_{4,1}={}&
\frac{315(1-e^2)}{64 e^{10}}\bigg(
1-\frac{e^2}{3}-\frac{2e^4}{15}-\frac{8e^6}{105}-\frac{16 e^8}{315}
\nonumber
\\
\quad
&-\frac{\sqrt{1-e^2}}{e}\arcsin e
\bigg).
\end{align}


\vspace*{\fill}  
\begin{flushright}
\begin{minipage}{0.45\textwidth}
\subsection{Prolate case ($a<b$)}
We list the explicit forms of $C_{\alpha, 0}$ (for $\alpha=0, 1, 2, 3, 4, 5$) 
and $C_{\alpha, 1}$ (for $\alpha=0, 1, 2, 3, 4$) in the prolate case as follows:
\begin{align}
C_{0,0}={}&
\frac{2}{e}\arctanh e,
\\
C_{1,0}={}&
\frac{1}{e^2}\left(
1-\frac{1-e^2}{e}\arctanh e
\right),
\\
C_{2,0}={}&
\frac{3}{4e^4}\left(
-1+\frac{5e^2}{3}+\frac{(1-e^2)^2}{e}\arctanh e
\right),
\\
C_{3,0}={}&
\frac{15}{24 e^6}\left(
1-\frac{8e^2}{3}+\frac{11 e^4}{5}
-\frac{(1-e^2)^3}{e}\arctanh e
\right),
\\
C_{4,0}={}&
\frac{35}{64 e^8}\bigg(
-1+\frac{11e^2}{3}-\frac{73 e^4}{15}+\frac{93 e^6}{35}
\nonumber
\\
&\quad
+\frac{(1-e^2)^4}{e}\arctanh e
\bigg),
\\
C_{5,0}={}&
\frac{63}{128 e^{10}}\bigg(
1-\frac{14 e^2}{3}+\frac{128 e^4}{15}-\frac{158 e^6}{21}+\frac{193 e^8}{63}
\nonumber
\\
&\quad 
-\frac{(1-e^2)^5}{e}\arctanh e
\bigg),
\end{align}
\begin{align}
C_{0,1}={}&
\frac{2}{e^2}\left(-1+\frac{\arctanh e}{e}\right),
\\
C_{1,1}={}&
\frac{3}{e^4}\left(1-\frac{2e^2}{3}-\frac{1-e^2}{e}\arctanh e\right),
\\
C_{2,1}={}&
\frac{15}{4e^6}\left(-1+\frac{5e^2}{3}-\frac{8e^4}{15}
+\frac{(1-e^2)^2}{e}\arctanh e
\right),
\\
C_{3,1}={}&
\frac{35}{8e^8}\bigg(
1-\frac{8e^2}{3}+\frac{11e^4}{5}-\frac{16e^6}{35}
\nonumber
\\
&\quad
-\frac{(1-e^2)^3}{e}\arctanh e\bigg),
\\
C_{4,1}={}&
\frac{315}{64 e^{10}}\bigg(
-1+\frac{11e^2}{3}-\frac{73 e^4}{15}+\frac{93 e^6}{35}-\frac{138 e^8}{315}
\nonumber
\\
&\quad +\frac{(1-e^2)^4}{e} \arctanh e\bigg).
\end{align}
\end{minipage}
\end{flushright}

\end{document}